**Auger-mediated radiative recombination in three-dimensional silicon/silicon-germanium nanostructures**

E.-K. Lee[*], D. J. Lockwood[†], J.-M. Baribeau[†], A. M. Bratkovsky[¥], T. I. Kamins[¥] and L. Tsybeskov[*]

[*]Department of Electrical and Computer Engineering, New Jersey Institute of Technology, Newark, NJ 07102, USA

[†]Institute for Microstructural Sciences, National Research Council, Ottawa, Canada, K1A 0R6

[¥]Hewlett-Packard Laboratories, Palo Alto, California 94304, USA

**In a semiconductor heterostructure with type II energy band alignment, the spatial separation between electrons and holes slows down their radiative recombination. With increasing excitation intensity, Auger recombination quickly becomes the dominate recombination channel, and it produces carrier ejection from the quantum well. Here, we show that in Si/SiGe three-dimensional nanostructures, this efficient process facilitates the formation of an electron-hole plasma (EHP) and/or electron-hole droplets (EHDs) in thin Si barriers separating SiGe clusters. In contrast to conventional, strongly temperature dependent and slow radiative carrier recombination in bulk Si, this EHD/EHP luminescence in nanometer-thick Si layers is found to be nearly temperature independent with radiative lifetime approaching $10^{-8}$ s, which is only slightly slower than that found in direct band gap III-V semiconductors.**

Photoluminescence (PL) in three-dimensional (3D) or cluster-like morphology Si/SiGe nanostructures is known to be efficient at low excitation intensity, but it quickly saturates as the excitation intensity increases[1-3]. An explanation involving Auger processes in Si/SiGe clusters has been proposed[4, 5], and calculations show that in a Si/SiGe quantum well having a thickness of several nanometers the Auger rate increases by ~100 times compared to that in bulk Si and Ge[6]. It has also been suggested that the suspected type II energy band alignment at the Si/SiGe hetero-interface[7] is responsible for the recently reported very long (up to $10^{-2}$ s) carrier radiative lifetime[5]. The combination of the enhanced Auger processes, very low density of structural defects[2] and extremely slow radiative recombination[5], makes a 3D Si/SiGe nanostructure an ideal "Auger fountain" emitter, i.e., a system where Auger recombination ejects carriers from the energy well into the energy barrier[8]. In this Letter, we show that in Si/SiGe 3D nanostructures this mechanism is responsible for efficient electron-hole accumulation within the Si barriers, and that this process radically modifies carrier recombination in the nanometer-thick Si layers.

The operational mode of the Auger fountain is revealed from excitation-dependent PL spectroscopy and time-resolved PL studies. Figure 1 shows PL spectra measured under different levels of photoexcitation. Using relatively low (~ 10 W/cm$^2$) continuous wave (CW) excitation, we observe a broad PL feature at 0.75-0.9 eV associated with SiGe clusters and a much weaker sharp PL band associated with Si band-to-band recombination at 1.096 eV[2]. On increasing the excitation intensity we find, in agreement with Refs. 2-4, a rapid saturation of the SiGe cluster PL. Under pulsed excitation with duration of ~ 6 ns and a peak intensity ≥ 1 kW cm$^2$, SiGe cluster PL becomes negligible compared to Si PL. At the same time, a broader PL peak at 1.079 eV appears. With the excitation intensity gradually increasing up to 1 MW/cm$^2$ (or energy density up to 10 mJ/cm$^2$), this broad PL peak shifts slightly to a higher photon energy (~ 1.085 eV), and quickly becomes the dominant PL feature (Fig. 1). We find that the PL intensity of this band as a function of excitation intensity follows the dependence $I_{PL} \sim I_{ecx}^n$, with exponent *n* in the range 1.2-1.8. This faster growing and broader PL band is associated with the recombination of electron-hole droplets (EHDs) in Si at low temperature and of an electron-hole plasma (EHP) at elevated temperatures[9-11].

Figure 2 compares the low-temperature dynamics in the PL bands associated with SiGe clusters and Si EHDs. The PL associated with SiGe clusters rises practically instantaneously (faster than the time resolution of our experiments, which is 2 ns), while the PL associated with Si EHDs has a rise time $\tau_{SiGe}^{rise} \geq 20$ ns. Under pulsed excitation with an energy density of 0.1mJ/cm$^2$, the SiGe cluster PL decay is non-exponential, with a fast initial decay (< 20 ns) followed by a much slower decaying (10$^{-4}$-10$^{-2}$ s) PL. At the same time, the Si EHD PL has a nearly single exponential decay with a characteristic lifetime $\tau_{EHD} \geq 50$ ns (Fig. 2a). Using a 100 times higher excitation intensity, we find an acceleration of the fast component of SiGe cluster PL decay surprisingly followed by a non-monotonic, at first rising and then falling, PL signal (Fig. 2b). Under the same excitation conditions, the Si EHD PL exhibits a nearly exponential decay with a lifetime of ~ 200 ns followed by a very fast (faster than 20 ns) decay (Fig. 2b).

The PL intensity at ~ 1.08 eV as a function of temperature (Fig. 3) shows a strong dependence on the excitation intensity. At low excitation with an energy density of ~ 100 μJ/cm$^2$, the PL intensity rapidly decreases as the temperature increases. Under the highest used excitation intensity of 10mJ/cm$^2$ the PL intensity is temperature independent up to nearly room temperature, as shown in Fig. 3.

The anti-correlation between the 0.75 -0.9 eV (SiGe cluster) and ~ 1.079-1.085 eV (Si EHD) PL band intensity behaviors, which is clearly depicted in Fig. 2b, suggests that the observed Si EHD PL mostly originates within the thin (~ 15 nm thick) Si layers separating the SiGe clusters rather than within the Si substrate. These details of our experimental results may be explained using a type II energy band alignment at the Si/SiGe cluster hetero-interfaces, and the required electronic transitions [steps (1) to (4)] are summarized schematically in Fig. 4.

(1) At low excitation intensity, the dominant relaxation process is slow radiative recombination between electrons localized in nanometer-thick Si layers and holes localized in SiGe clusters (i.e., indirect excitons). Due to a relatively weak wavefunction overlap between the spatially separated electrons and holes, this recombination at the Si/SiGe hetero-interface exhibits a long radiative lifetime $\tau_{SiGe}$ ranging from $10^{-4}$ to $10^{-2}$ s[5].

(2) At a high level of excitation, a fast non-radiative SiGe Auger recombination with a characteristic lifetime $\tau_{Auger} \leq 10^{-8}$ s successfully competes with the slow indirect exciton recombination at the Si/SiGe hetero-interface[4]. This process generates 'Auger holes' with energies of 0.7-0.8 eV, while the valence energy barrier at the Si/SiGe hetero-interface is ~ 0.3-0.4 eV high. Thus, the Auger hole ejection from SiGe clusters into Si barriers is quite efficient.

(3) The efficient Auger-mediated hole transfer from SiGe clusters into Si barriers produces electron-hole condensation in the Si barriers, and the EHD lifetime is approaching ~ 50 ns (Fig. 1, 2). On increasing the excitation intensity, we find that the Si EHD PL lifetime decreases while the PL intensity increases. Thus, we think that the observed fast PL decay is mainly associated with a Si EHD short radiative lifetime rather than with a competing non-radiative recombination channel. In agreement with this conclusion, we do not find Si EHD PL intensity saturation as a function of excitation intensities up to ~ 10 mJ/cm$^2$.

(4) In Si, the EHD condensation, which is a first order phase transition, depends on the excess carrier concentration and temperature[11,12]. According to our data (Fig. 2b), it takes ~ $10^{-6}$ s after the excitation pulse for the carrier concentration to decrease (because of recombination) below the threshold and for the EHD/electron-hole-gas phase transition to take place. Thus, a hole, which is no longer confined within an EHD, could be recaptured by a partially emptied SiGe cluster, and an increased electron-hole pair concentration at the Si/SiGe hetero-interface would result in the observed increase of the PL intensity at ~ 0.82 eV.

The proposed luminescence mechanism is qualitatively similar to the previously reported "Auger fountain" in III-V heterostructures with type II energy band alignment[8].

However, in 3D Si/SiGe nanostructures, Auger-excited holes and photo-generated electrons recombine in the nanometer-thick Si barriers, and Si is known to be the classic example of an indirect band gap semiconductor with slow carrier radiative recombination and low PL quantum efficiency, especially at room temperature[13]. Why, then, can we observe an efficient luminescence with nanosecond lifetime and with a practically negligible intensity temperature dependence up to room temperature?

Our explanation is based on the qualitative difference between recombination conditions in the nanometer thick Si layer under a low and high level of photoexcitation. At low excitation, the Si free exciton radiative lifetime is $\tau_{Si}^{radiative} \sim 10^{-3}$ s [13], mainly due to the low probability of a three particle (electron-hole-phonon) process. As the temperature increases, the PL intensity decreases exponentially, with a thermal quenching activation energy equal to the exciton binding energy[2]. Thus, at elevated temperatures, charge carriers are no longer bound together, and it is faster for electrons and holes to find structural and/or interface defects and quickly recombine non-radiatively. In the simplest model, the low carrier radiative recombination rate does not change as function of temperature but the non-radiative recombination rate increases exponentially as the temperature increases[5].

At high excitation intensity, especially when the excess carrier concentration is approaching the density-of-states (DOS) ($\sim 10^{19}$ cm$^{-3}$ in bulk Si[14]), the recombination conditions change drastically. At low temperatures, the main radiative energy dissipation channel is the recombination of EHDs. In bulk Si, EHDs are produced directly from the hot, dense photo-excited EHP[11], while in 3D Si/SiGe nanostructures their formation is controlled by the Auger-mediated supply of hot holes ejected from SiGe clusters (Fig. 4). Compared to the free exciton, the EHD PL peak photon energy is reduced due to strong carrier-carrier interactions[15]. Also, the electron-phonon interaction increases the local lattice temperature and produces additional non-equilibrium phonons with an energy-momentum dispersion quite different compared to that at thermal equilibrium[16-18]. Thus, the significantly disturbed phonon spectrum increases the probability of phonon-assisted transitions, and the recombination lifetime drops down to $10^{-7}$-$10^{-8}$ s (Fig. 2). Under these conditions, the light emission could be quite efficient: an estimate of the upper limit for the EHD light emission internal quantum efficiency $\eta$ can be made by comparing exciton and EHD lifetimes and using $\eta = \tau_{rad}^{-1}/(\tau_{rad}^{-1} + \tau_{non\text{-}rad}^{-1})$, where $\eta = \eta_{exc} \approx 10^{-4} - 10^{-6}$ for exciton recombination[19]. Assuming that $\tau_{non\text{-}rad}$ remains the same, the highest possible EHD light emission quantum efficiency can easily reach large values, such as $\eta = \eta_{EHD} \sim 10^{-1} - 10^{-2}$. Also, at these excitation intensities, carrier diffusion along the nanometer thick Si layers is strongly suppressed since the carrier concentration becomes comparable to the DOS. Thus, a high concentration of excess carriers provides an efficient (because of fast transitions) and nearly temperature independent (because of the suppressed diffusion and, possibly, saturation of non-radiative recombination channels) radiative recombination. Also, our measurements indicate that as the temperature increases and a transition from EHD to EHP takes place, the carrier radiative lifetime remains approximately the same (not shown).

Note that at a very high carrier concentration, non-radiative band-to-band Auger recombination eventually should start competing with Si EHD/EHP recombination. Auger recombination has a lifetime $\tau_{Auger} = \dfrac{1}{Cn^2}$, where $n$ is the carrier concentration, and for bulk silicon $C \approx 10^{-30(-31)}$ cm$^6$s$^{-1}$ for electrons (holes), respectively[14]. Thus, our data confirms the existence of a "window of opportunity": at a carrier concentration high enough to form EHD/EHP, which is estimated to be $\sim 10^{18}$ cm$^{-3}$, the shortest experimentally detected recombination lifetime is found to be $\sim 5 \cdot 10^{-8}$ s. This fast, presumably radiative, recombination channel successfully competes with the non-radiative band-to-band Auger process, which has a longer ($\sim 10^{-6}$ s) characteristic lifetime. This "window" in 3D Si/SiGe nanostructures seems to be significantly expanded by a reduction of the DOS and the suppression of carrier diffusion in the nanometer-thick partially-strained Si layers sandwiched between Ge-rich SiGe clusters[20].

Finally, we address the long-standing issue of the feasibility of Si-based light emitters and lasers. Compared to other approaches (see Refs. 21, 22), the carrier recombination mechanism demonstrated here exhibits a short and practically temperature independent radiative lifetime, and that makes this particular route an extremely promising one for CMOS-compatible monolithically-integrated optoelectronic systems.

**Acknowledgement**

We thank Dr. X. Wu for the TEM micrograph shown in Fig. 1. The work at the New Jersey Institute of Technology has been supported in part by Intel, Semiconductor Research Corporation, Foundation at NJIT and US National Science Foundation.

**Methods**

The samples used in this study were grown by MBE in a VG Semicon V80 system. The structures, grown on (001) Si at temperature $T_G$= 650 °C, consist of 10 period Si/Si$_{1-x}$Ge$_x$ multiple layers with $x \approx 0.55$ close to the middle of the SiGe cluster (Fig. 1, inset). For details of the sample preparation, see Ref. 20. Transmission electron microscopy studies have shown that the Si/Si$_{1-x}$Ge$_x$ multilayers exhibit an island-like morphology (i.e., 3D growth) embedded into a Si matrix[20]. The Si$_{1-x}$Ge$_x$ island height and Si separating layer thickness were kept constant throughout the entire multilayer structure.

The PL measurements were performed using a closed-cycle vacuum cryostat in the temperature range of 15 – 300 K. For CW PL measurements, we used an argon ion laser (476.5 nm) with the excitation intensity varied from 0.1 to 10 W/cm$^2$. The PL signal was dispersed using a single grating Acton Research 0.5 m monochromator and detected by a cooled Hamamatsu photomultiplier in the spectral range 0.9 – 1.6 μm (0.77 – 1.38 eV). For PL measurements under pulsed laser excitation, the second harmonic of a Nd:YAG laser was used with a wavelength of 532 nm, a 6 ns pulse duration, and a known pulse energy density. The PL decay signal was stored in a LeCroy digital storage oscilloscope. The overall time resolution of the entire system was $\geq 2$ ns.

**Figure captions:**

**Figure 1.** The PL spectra in 3D Si/SiGe nanostructures measured under different excitation conditions (shifted vertically for clarity). The PL peaks associated with SiGe clusters and free-excitons (FE) and electron-hole droplets (EHD) in Si are indicated. The inset shows a dark-field cross-sectional transmission electron micrograph (TEM) with SiGe clusters as lighter areas separated by ~ 10-15 nm thick Si layers.

**Figure 2.** The PL dynamics under pulsed laser energy densities of (a) 0.1 and (b) 10 mJ/cm$^2$ recorded at photon energies associated with SiGe cluster PL (~ 0.82 eV) and Si EHD PL (~ 1.085 eV).

**Figure 3.** The PL peak intensity at ~ 1.08 eV (Si EHD) as a function of reciprocal temperature measured at different excitation energy densities, as indicated.

**Figure 4.** Schematic representation of different recombination processes in Si/SiGe nanostructures with a type II energy band alignment. At very low excess carrier concentration, the slow SiGe recombination is due to an indirect (in real space) exciton formed by an electron localized in Si and a hole localized in SiGe (process 1). At high excess carrier concentration, Auger recombination ejects hot holes from the SiGe wells (i.e., valence energy wells) directly into the Si barriers (2) and facilitates EHD formation/recombination within the Si barriers (3). During ~ 0.1- 1 μs after the laser pulse, fast EHD recombination strongly reduces the excess carrier concentration. Then, holes are recaptured by the emptied SiGe energy wells (4) and recombine with electrons localized in the Si barriers (1).

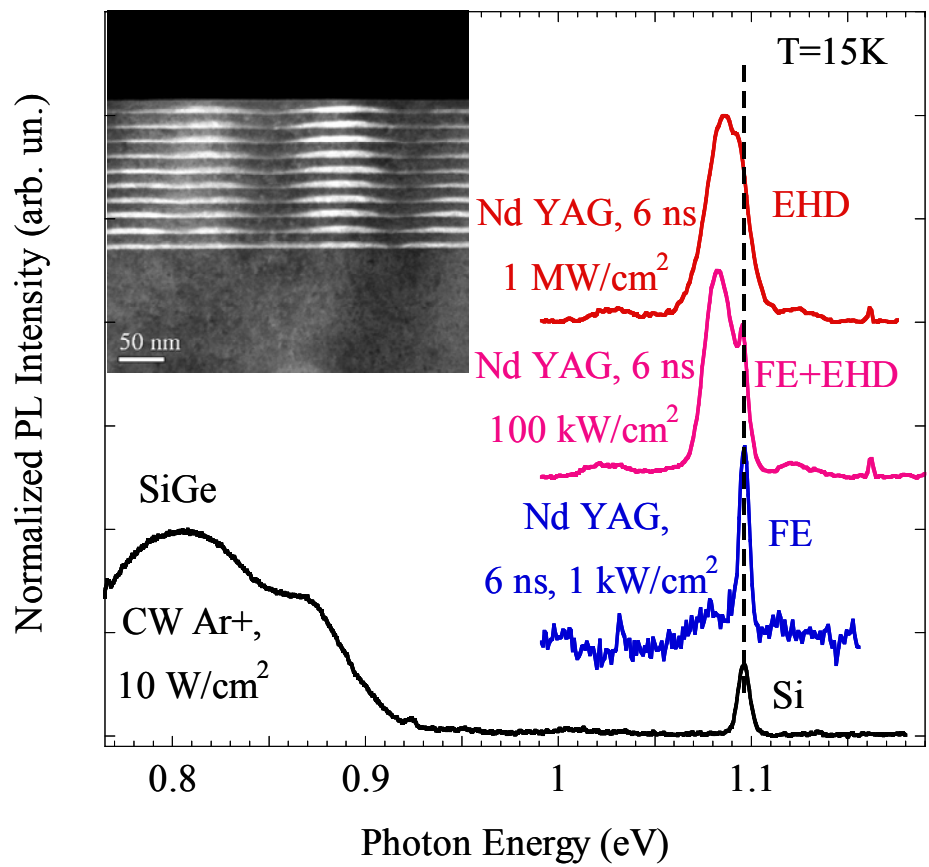

Lee et al., Fig. 1.

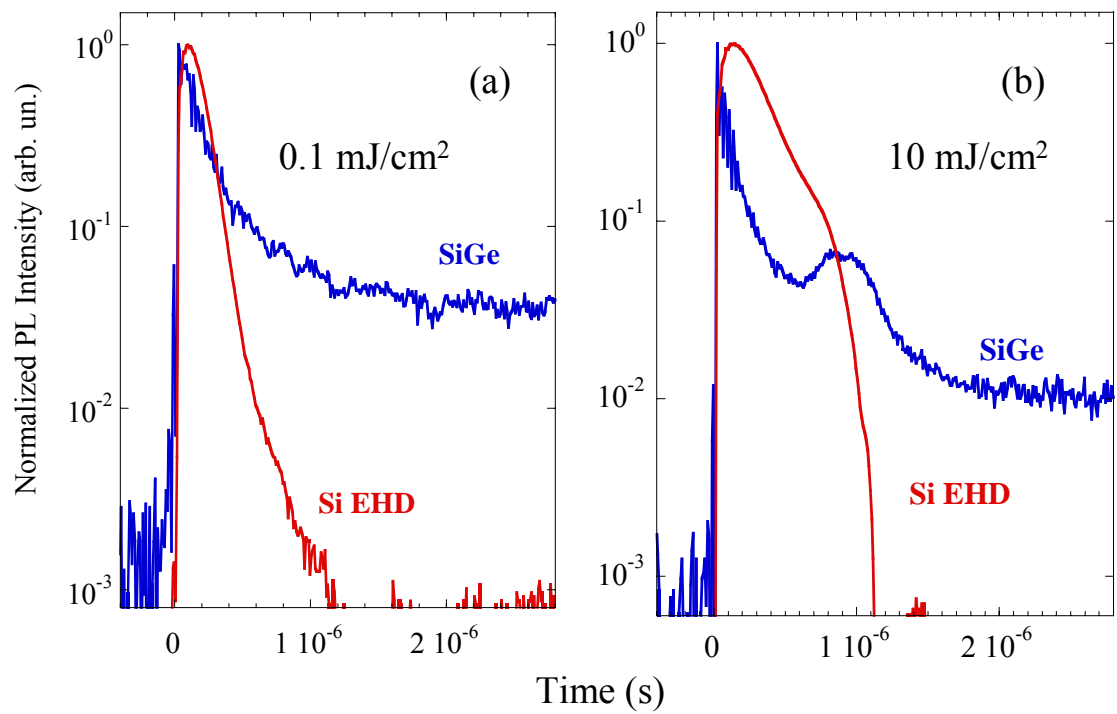

Lee et al., Fig. 2.

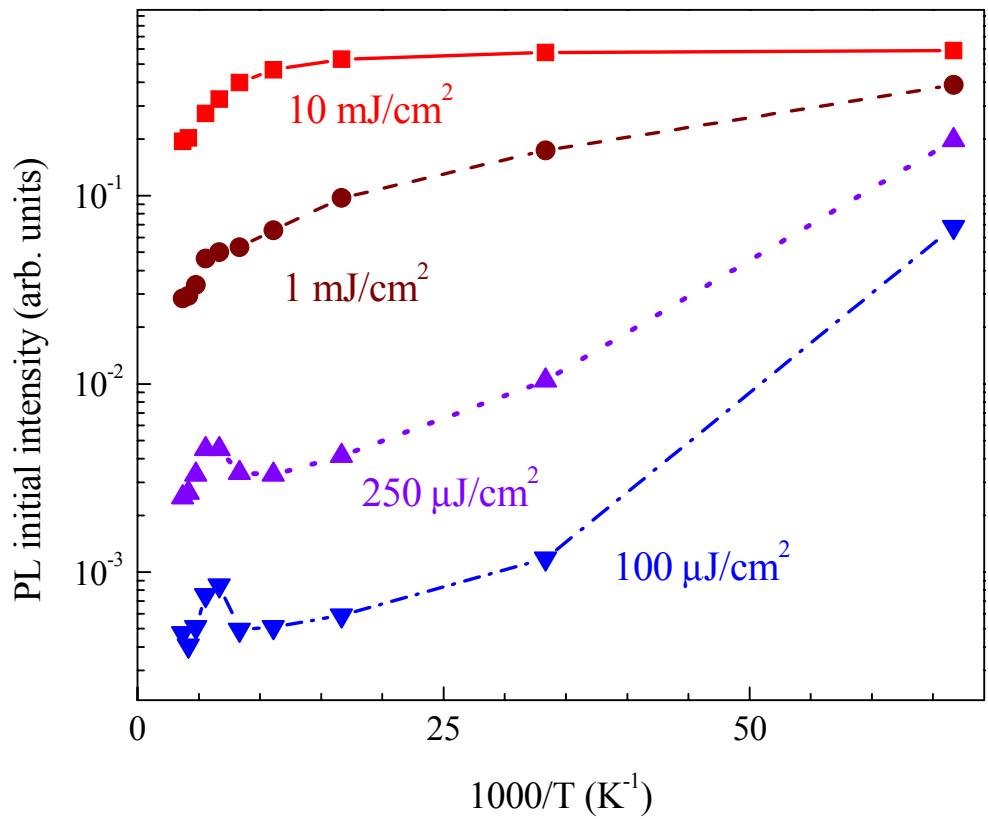

Lee et al., Fig. 3.

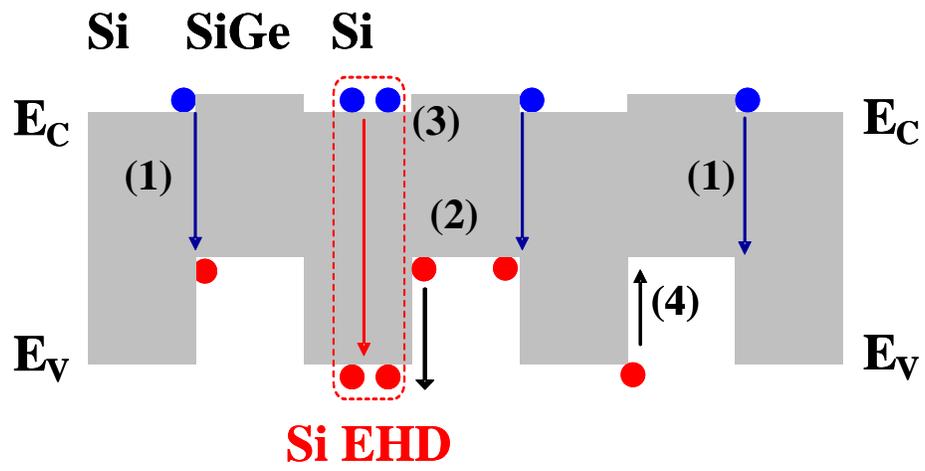

Lee et al., Fig. 4.